\documentclass[useAMS,usenatbib]{mn2e}

\usepackage[dvips]{graphicx}
\usepackage{wrapfig}
\usepackage{color}
\newcommand{\lseq}{\mbox{\raisebox{-0.7ex}{$\;\stackrel{<}{\sim}\;$}}}
\newcommand{\gseq}{\mbox{\raisebox{-0.7ex}{$\;\stackrel{>}{\sim}\;$}}}

\title[Escape of ionizing photons]
{The escape of ionizing photons from supernova-dominated primordial galaxies}
\author[Yajima et al.]
{Hidenobu Yajima$^{1}$\thanks{E-mail:yajima@ccs.tsukuba.ac.jp(HY); 
umemura@ccs.tsukuba.ac.jp(MU); mmori@ccs.tsukuba.ac.jp(MM); 
nakamoto@geo.titech.ac.jp(TN)}, 
Masayuki Umemura$^{1}$, 
Masao Mori$^{1}$, 
and Taishi Nakamoto$^{2}$
\\
$^{1}$Center for Computational Sciences, University of Tsukuba, 
Tsukuba 305-8577, Japan\\
$^{2}$Earth and Planetary Sciences, Tokyo Institute of Technology, 2-12-1 Ookayama, Meguro-ku,
Tokyo, 152-8551, Japan
}
\begin{document}

\date{Accepted ?; Received ??; in original form ???}

\pagerange{\pageref{firstpage}--\pageref{lastpage}} \pubyear{2008}

\maketitle

\label{firstpage}

%----------------------------------------------------------------------
%
% Abstract
%
%----------------------------------------------------------------------
\begin{abstract}

In order to assess the contribution of Lyman break galaxies (LBGs) 
and Lyman alpha emitters (LAEs) at redshifts 
$3<z<7$ to the ionization of intergalactic medium (IGM),
we investigate the escape fractions of ionizing photons from 
supernova-dominated primordial galaxies by solving 
the three-dimensional radiative transfer. 
%\textcolor{blue}{
The model galaxy is employed from an ultra-high-resolution chemodynamic 
simulation of a primordial galaxy by Mori \& Umemura (2006),
which well reproduces the observed properties of LAEs and LBGs.
The total mass of model galaxy is $10^{11}M_\odot$.
We solve not only photo-ionization but also collisional ionization by shocks.
In addition, according to the chemical enrichment, we incorporate 
the effect of dust extinction, taking the size distributions of dust into account. 
As a result, we find that dust extinction reduces the escape fractions by a factor $1.5-8.5$
in the LAE phase and by a factor $2.5-11$ in the LBG phase, while the collisional
ionization by shocks increases the escape fractions by a factor $\approx 2$.
The resultant escape fractions are $0.07-0.47$ in the LAE phase and
$0.06-0.17$ in the LBG phase. These results are well concordant with 
the recent estimations derived from the flux ratio at 
1500~$\rm \AA$ to 900~$\rm \AA$ of LAEs and LBGs.
%}
Combining the resultant escape fractions with the luminosity functions 
of LAEs and LBGs, we find that high-$z$ LAEs and LBGs can 
ionize the IGM at $z=3-5$. 
However, ionizing radiation from LAEs as well as LBGs falls short to ionize 
the IGM at $z>6$. That implies that additional ionization sources may required 
at $z>6$. 
\end{abstract}

%----------------------------------------------------------------------
%
% Keywords
%
%----------------------------------------------------------------------

%----------------------------------------------------------------------
%
% Section 1: Introduction
%
%----------------------------------------------------------------------
\section{INTRODUCTION}
One of momentous issues regarding the evolution of intergalactic 
medium (IGM) is the ionization history of the universe,
which significantly influences the galaxy formation 
\citep[e.g.,][]{Susa00,Umemura01,Susa04}. 
The Wilkinson Microwave Anisotropy Probe (WMAP) results provide 
a wealth of information about the cosmic reionization \citep{Pa07, Du09}.
However, the detailed history of reionization 
and the nature of ionizing sources are not yet fully understood.
\citet{Haa96} pointed out that the UV background radiation is 
dominated by quasars at $z<4$. 
\citet{Fan01} showed, using the SDSS sample, that the bright-end slope 
of the quasar luminosity function at $z\gseq4$ are considerably shallower 
than that at low-redshifts, 
and they concluded that quasars cannot maintain the ionization 
of IGM at $z\gseq4$. 
Subsequently, a lot of arguments have been concentrated on the 
possibility that the IGM is ionized mainly by 
UV radiation from high-$z$ star-forming
galaxies like Lyman break galaxies (LBGs) and Lyman $\alpha$ emitters (LAEs)
\citep[e.g.,][]{Fan06,Bou07,Gne08b}. 
However, the estimate of the contribution of LBGs or LAEs suffers
significantly from the ambiguity regarding the escape fractions of
ionizing photons from star-forming galaxies \citep{Raz06, GKC08}. 

Observationally, the escape fractions of ionizing photons are assessed by
the relative escape fractions $f_{\rm esc,rel}$, 
using the flux ratio at 1500~$\rm \AA$ to 900~$\rm \AA$, $(F1500/F900)_{\rm obs}$,
which are defined by
\begin{equation}
f_{\rm esc,rel} = \frac{(L1500/L900)_{\rm int}}{(F1500/F900)_{\rm obs}} 
\exp(\tau^{\rm IGM}_{900}), \label{f_esc_rel}
\end{equation}
where $\tau^{\rm IGM}_{900}$ 
represents the line-of-sight opacity of the IGM for 900~$\rm \AA$ photons. 
Normally, the intrinsic luminosity ratio at 
1500~$\rm \AA$ to 900~$\rm \AA$, $(L1500/L900)_{\rm int}$, is
assumed to be 3 as a fiducial value.
\citet{St01} found $f_{\rm esc,rel} \gseq  0.5 $ from the 
composite spectrum of 29 LBGs at $z\sim3$, and \citet{Gi02} 
and \citet{In05} estimated the upper limit of $f_{\rm esc,rel} \lseq 0.1-0.4$ 
for some LBGs at $z \sim 3$.
The direct detection of ionizing photons from high-$z$ star-forming 
galaxies has been accomplished recently as a consequence of intensive and 
continuous exertion. \citet{Sha06} detected the escaping ionizing 
photons from 2 LBGs in the SSA22 field at $z=3.1$ and they estimated the 
average relative escape fraction $f_{\rm esc,rel}= 0.14$. 
%----------------------------------------------------------
%HY
%More recently, \citet{Iwa09} successfully discovered the Lyman 
%continuum emission from the total number of 197 LAEs and LBGs in SSA22 
%field. 
Moreover, \citet{Iwa09} successfully detected the Lyman continuum emission
 from 10 LAEs and 7 LBGs within 197 samples of LAEs 
 and LBGs in the SSA22 field.
%They derived that the relative escape fractions of these young 
%galaxies are greater than 0.15.
%
%HY(revised)
%\textcolor{red}{
They have shown that the mean value of relative escape fractions 
for 7 LBGs is 0.11 after a correction for dust extinction, and
can be 0.20 if IGM extinction is taken into account.
%}

%----------------------------------------------------------
%HY(new)!
%So far three-dimensional radiation transfer calculations always predict 
%one order of magnitude smaller escape fraction (a few \%) than
%the observations quoted above \citep{Raz06, Raz07, Gne08b}. 
Theoretically, the accurate estimation of escape fractions requires
the three-dimensional (3D) radiative transfer calculations of ionizing
photons traveling in inhomogeneous interstellar medium. 
Such 3D radiative transfer calculations are fairly recent attempts,
because a great deal of computations are demanded. 
First, \citet{Cia02} calculated the escape fractions in simple model clouds, 
where the density distributions are smoothed Gaussian or fractally inhomogeneous.
More recently, using gasdynamical simulations of galaxy formation,
the escape fractions are estimated by solving 3D radiative transfer.
\citet{Raz06,Raz07} have found that the escape fractions decline
from several per cent at $z=3.6$ to 0.01-0.02 at $z=2.39$, 
due to higher gas clumping at lower redshifts. 
But, in these simulations, the effects of dust extinction are not
taken into account. 
%----------------------------------------------------------
\citet{Gne08a} performed a cosmological simulation on the formation of 
a disk-like galaxy that provides the three-dimensional distributions of 
absorbing gas in and around the galaxy.
%where the star formation rate is relatively
%low as $\lseq$ a few $M_\odot$ yr$^{-1}$. 
Then, 3D radiative transfer was solved
including the dust extinction.
As a result, the escape fractions turned out to be as low as a few per cent.
In this model, the bulk of stars is embedded deep inside 
the optically-thick HI disk and therefore most of ionizing photons 
emitted from hot stars are absorbed by neutral hydrogen in the HI disk.
Only a small fraction of ionizing photons from the stars
that are located near the edge of disk can escape from the galaxy. 
Hence, almost regardless of the effects of dust extinction,
the resultant escape fractions become quite small.
%  MM
%\textcolor{magenta}{
%All of these simulations always predict one order of magnitude smaller
%escape fraction than that of the observational estimation.
%}
It implies that star-forming galaxies cannot give a significant contribution 
to the IGM ionization at $z\gseq 3$.
%\textcolor{magenta}{
For higher redshift $z \gseq 8$, on the other hand, some numerical
simulations have already shown that large escape fractions of some tens
per cent are possible for low-mass galaxies, which include Pop III stars
\citep{ABS06, Wh04, Ki04, WC09}.
%}
%%%Such small escape fractions
%%%are one order of magnitude smaller than the observational estimates. 
%%%It implies that star-forming galaxies cannot give a significant contribution 
%%%to the IGM ionization at $z\gseq 3$. 
%HY(revised)
%\textcolor{red}{
%Very recently, \citet{WC09} performed 3D radiation hydrodynamic simulations
%to assess the contribution of dwarf galaxies to cosmic reionization 
%at redshift $z=8$. They studied the UV escape fractions for 
%low-mass galaxies in the mass range of 
%$M_{\rm total}=3\times10^{6}-3\times10^{9}M_{\odot}$.
%As a result, they have shown that the UV escape fractions can reach
%up to $\sim$ 0.8 without dust extinction 
%in halos with $>10^8M_\odot$ for a top-heavy initial mass function (IMF).
However, in order to assess the contribution of
LAEs and LBGs observed at $3 \lseq z \lseq 7$
to IGM reionization, we should evaluate the UV escape fractions from more massive galaxies.
The escape of UV photons can sensitively depend on the gravitational
potential (Whalen et al. 2004, Kitayama et al. 2004). 
Also, dust extinction should be taken into account in such primordial galaxies.
%}

In this paper, we reconsider the escape fractions 
of ionizing photons from high-$z$ primordial galaxies
by employing a supernova-dominated primordial galaxy model proposed by 
\citet{MU06}, which 
well reproduces the observed properties of LAEs and LBGs.
%\textcolor{red}{
The total mass is $10^{11}M_\odot$, and the star formation
with the Salpeter IMF is included. 
%}
In this model, stars are distributed more extendedly in the galaxy and
the bulk of interstellar gas is collisionally ionized by supernova-driven shock.
The spread of stellar distributions and the shock heating
can lead to diminishing the absorption of ionizing photons 
and may enhance escape fractions. 
Here, we perform the 3D radiative transfer calculations including not only
the photo-heating but also the shock heating. 
We incorporate the dust extinction that is consistent 
with the chemical evolution of galaxy. Then, 
using the resultant escape fractions, we explore 
whether LAEs and LBGs can contribute to the ionization 
of IGM in a high-$z$ universe. In the present analyses,
the cosmological parameters are assumed to be 
$H_{\rm 0}=70$ km s$^{-1}$ Mpc$^{-1}$, 
$\Omega_{\rm M}=0.3$ and $\Omega_{\rm \Lambda}=0.7$.
In \S 2, the model and numerical method are described.
In \S 3, the numerical results on escape fractions are provided.
In \S 4, we argue the contributions of LAEs and LBGs to the IGM ionization. 
In \S 5 is devoted to the summary. 

%----------------------------------------------------------------------
%
% Section 2:  Model and Method
%
%----------------------------------------------------------------------
\section[]{MODEL AND METHOD}
As a model galaxy, we adopt the high-resolution hydrodynamic 
simulations ($1024^3$ fixed Cartesian grids) by \citet{MU06},
which are coupled with the collisionless dynamics for dark matter 
particles as well as star particles and also the chemical enrichment.
The simulation box size is 134 kpc in physical scales.
\citet{MU06} demonstrated that an early proto-galactic evolution with
multitudinous type II SNe (SNeII) exhibits intense Lyman $\alpha$ 
emission, well resembling LAEs. Subsequently, the galaxy 
shifts to a stellar continuum radiation-dominated phase, which appears 
like LBGs.
%\textcolor{blue}{
In the present analysis, the stages from $t_{\rm age}=0.1$Gyr 
to $t_{\rm age}=0.3$ Gyr are defined
to be the LAE phase, and the stages from $t_{\rm age} = 0.5$ Gyr 
to $t_{\rm age}=1.0$ Gyr are
the LBG phase.
%}

As a post process, we calculate the escape fractions of ionizing photons 
with an accurate radiation transfer scheme.
The data of the hydrodynamic simulations are coarse-grained 
into $128^3$ Cartesian grids to solve radiation transfer.
We use the Authentic Radiation Transfer method (ART) developed by 
\citet{NUS01}. 
The performance of this scheme has already reported as a part 
of the comparison study by \citet{Il06}.
The radiation transfer equation is solved along $128^{2}$ rays 
with uniform angular resolution from each source.
At a point of optical depth $\tau$, the specific intensity is given
by $I_{\nu}(\tau) = I_{\nu}(0)\exp{(-\tau)}$,
where $I_{\nu}(0)$ is the intrinsic intensity of ionizing 
radiation and $\tau$ is the optical depth of neutral hydrogen and dust. 
As for scattering photons, we employ the on-the-spot approximation 
\citep{Os1989}, in which scattering photons are assumed to be absorbed 
immediately on the spot. 
We obtain the ionization structure assuming the ionization equilibrium,
$\Gamma^{\gamma}n_{\rm HI} + \Gamma^{C}n_{\rm HI}n_{\rm e} 
= \alpha_{\rm B}n_{\rm p}n_{\rm e} $,
where $\Gamma^{\gamma}, \Gamma^{C}$ and $\alpha_{\rm B}$ 
are the photo-ionization rate, the collisional ionization rate 
and the recombination rate to all excited states, respectively.
We continue the radiative transfer calculation recurrently 
until the ionization structure converges. 
%HY(revised)
%\textcolor{red}{
In a supernova-dominated model galaxy, the collisional ionization 
occurs mostly in low density, high temperature regions 
with $n \lseq 10^{-3}{\rm cm}^{-3}$ and $T_{\rm coll}>\rm{10^{4} K}$.
In such regions, stellar UV radiation contributes to the increase of 
ionization degree, but does not much to the increase of the temperature.
On the other hand, neutral regions are photoionized
and heated up to $\rm{\sim 10^{4} K}$ \citep{UI84,TW96}, when
ionizing radiation is irradiated.
Hence, in the present analysis, we assume the gas temperature 
to be $T=\max \{ T_{\rm coll},\rm{10^{4} K} \}$ in ionized regions, and
we do not update the temperature in iteration of radiation transfer calculation. 
To evaluate the escape fraction of ionizing photons,
we count all photons above the Lyman limit that escape from the calculation box. 
%}

The number of ionizing photons emitted from source stars 
is computed based on the theoretical spectral energy distribution (SED) 
given by a population synthesis scheme, P\'{E}GASE v2.0 \citep{Fi1997}. 
We assume the Salpeter (1955) initial mass function 
in the mass range of $0.1 M_{\odot} - 50 M_{\odot}$.
%HY(revised)
As for dust grains, we adopt the empirical size distribution 
$n_{\rm d}(a_{\rm d}) \propto a^{-3.5}_{\rm d}$ \citep{M77} 
in the range from 0.1$\rm \mu m$-1.0$ \rm \mu m$
, where $a_{\rm d}$ is the radius of a dust grain.
We assume refractory grains like silicates, for which
the photodestruction of dust by UV radiation is negligible
over the Hubble time-scale \citep{DS79}.
%----------------------------------------------------------
%HY
The dust grains are distributed proportionally to the metallicity calculated
in the hydrodynamic simulations with the relation of $m_{d}=0.01m_{g}(Z/Z_{\odot})$,
where $m_{d}$, $m_{g}$, and $Z$ are the dust mass, gas mass, and metallicity in a grid.
The density in a dust grain is assumed to be 3 g cm$^{-3}$ like silicates.
The dust opacity is given by 
%\begin{equation}
%d\tau _{\rm dust} = Q(\nu) \pi a^{2}_{\rm d} n_{\rm d} ds,
%\end{equation}
$d\tau _{\rm dust} = Q(\nu) \pi a^{2}_{\rm d} n_{\rm d} ds $,
where $Q(\nu), a_{\rm d}$ and $n_{\rm d}$ are the absorption $Q$-value, 
dust size and number density of dust grains, respectively.
Since the assumed range of dust size is larger than the wavelength of Lyman limit,
we assume $Q(\nu)=1$ for ionizing photons \citep{DL84}.

%----------------------------------------------------------------------
%
% Section 3:  Results
%
%----------------------------------------------------------------------
\section[]{RESULTS}

\subsection{Galactic Evolution}

\begin{figure*}
\begin{center}
\includegraphics[scale=1.0]{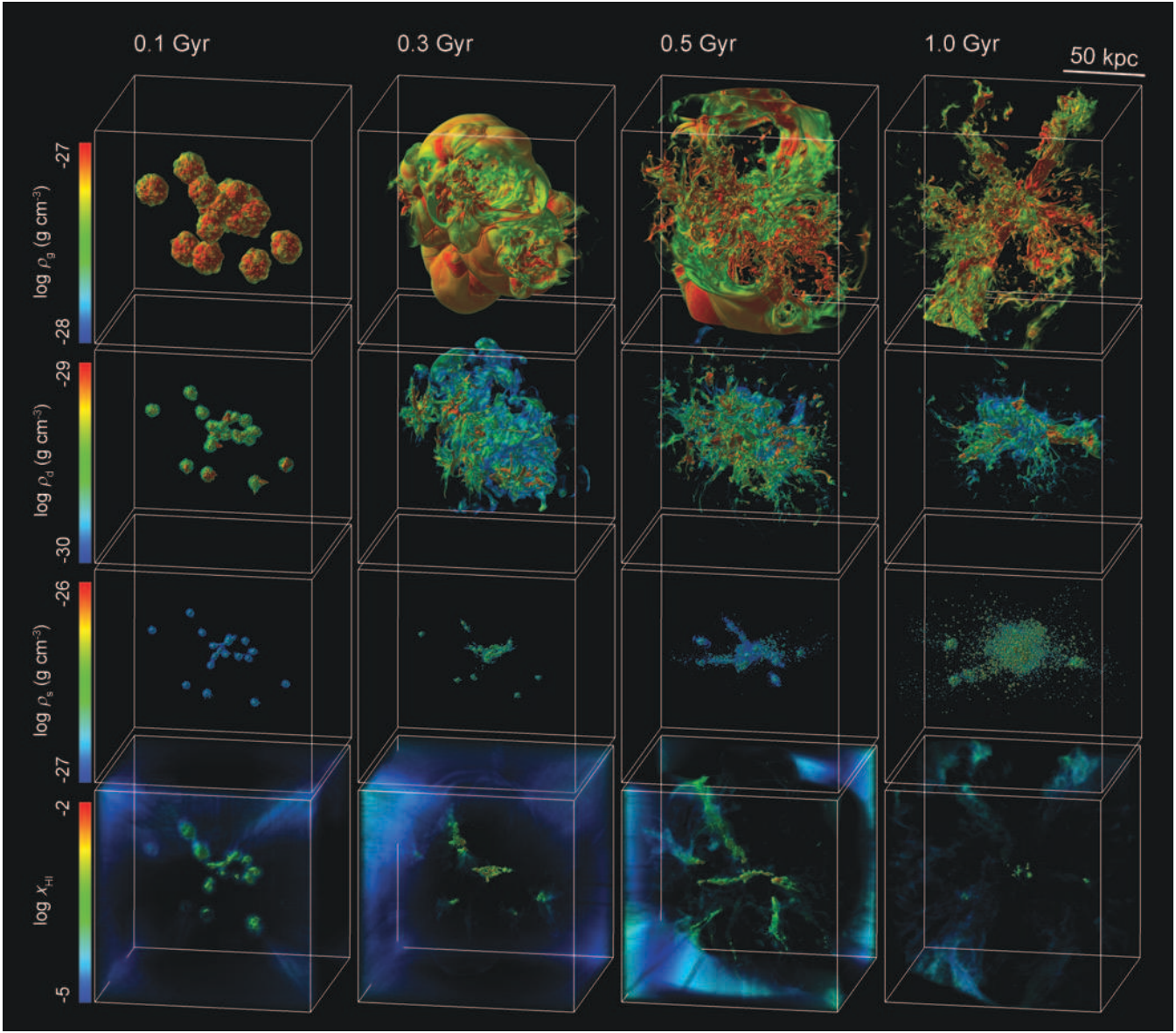}
\caption{
Snapshots of the evolution of the model galaxy 
with total mass of $10^{11} M_\odot$,
at $t_{\rm age}=0.1, 0.3, 0.5$ 
and 1.0 Gyr. Each panel in row corresponds the spatial distributions of gas 
density $\rho_{\rm g}$ (g cm$^{-3}$), dust density $\rho_{\rm d}$ 
(g cm$^{-3}$), stellar density $\rho_{\rm s}$ (g cm$^{-3}$), and 
fractions of neutral hydrogen in logarithmic scale $\chi_{\rm HI}$, 
respectively. The simulation box is 134 kpc in physical scale. 
}
\end{center}
\end{figure*}

Figure 1 shows the snapshots for the evolution of model galaxy 
as a function of redshift from $t_{\rm age}=0.1$ to 1.0 Gyr. 
The density of gaseous and stellar components are adopted from
\citet{MU06} . The dust density is evaluated by the metallicity. 
As a result of multiple supernova explosions, 
dust is distributed more extendedly than stars, which is significantly
relevant to the absorption of ionizing photons. 
Bottom panels show the calculated ionization structure in terms of
neutral hydrogen fractions $\chi_{\rm HI}$.

First, we see the evolution of the model galaxy.
At $t_{\rm age}= 0.1$ Gyr, stars form in high-density peaks in 
sub-galactic condensations and the burst of star formation starts. 
Then, massive stars in the star-forming regions explode as SNeII 
one after another. The gas in the vicinity of SNeII is quickly enriched 
with the ejected heavy elements and then interstellar dust 
is locally procreated. However, a large amount of gas still retains 
a metal-free and dust-free state. 
The spatial distributions of heavy elements are highly inhomogeneous, 
where gas enriched with $-5.1 \lseq {\rm [Si/H]} \lseq -1.1$ and 
$ -5 \lseq {\rm [O/H]} \lseq -1.0$ coexist with virtually metal-free 
gas. Supernova-driven shocks collide with each other to generate 
large-scale hot ($\geq 10^6$ K) bubbles reaching a higher ionization 
degree. At $t_{\rm age}=0.3$ Gyr, roughly 50 per cent of the total volume is highly ionized
with $\chi_{\rm HI} = 10^{-6} \sim 10^{-7}$, where the ionization degree is 
controlled by the collisional ionization by shock heating and the 
photo-ionization by UV radiation from hot young stars. 
The spatial distributions of the heavy elements are still highly 
inhomogeneous in the range of $-2.4 \lseq {\rm [Si/H]} \lseq -0.4$ and $-2.5 \lseq 
{\rm [O/H]} \lseq -0.5$. 

After $t_{\rm age}=0.5$ Gyr, the hot bubbles expand and 
blow out into the intergalactic space.
As a result, more than 80 per cent of volume is occupied with
highly ionized gas with $\chi_{\rm HI} = 10^{-6}\sim 10^{-7}$.
Finally, the merger of sub-galactic condensations promotes the mixing 
of heavy-elements and weakens the spatial inhomogeneities of heavy-element 
abundance and ionization degree.
As a result, the heavy-element abundance of interstellar medium (ISM) 
converges to 
$-0.4 \lseq {\rm [Si/H]} \lseq 0.1$ and $-0.3 \lseq {\rm [O/H]} \lseq 0.2$ 
with small dispersion, and eventually 95 per cent of volume is filled with
ionized gas. 
%\textcolor{red}{
This means that the metallicity reaches around the solar abundance
in $10^9$yr. It is consistent with the previous works on the elliptical
galaxy formation \citep{AY87,Gibson97,KA97,Kawata-etal03,KG03}.
The mean value of mass weighted heavy-element abundance
is ${\rm [Si/H]} \simeq -1.0$ and ${\rm [O/H]} \simeq -0.9$ at the LAE phase, 
while ${\rm [Si/H]} \simeq -0.4$ and ${\rm [O/H]} \simeq -0.3$ at the LBG phase.
These are translated into the mass weighted metallicity 
as 0.14 $Z_{\odot}$ at LAE phase
and 0.52 $Z_{\odot}$ at LBG phase, which are concordant 
with the observations by \citet{Pe01} and \citet{Ma09}. 
%}

%The luminosity of Lyman $\alpha$ emission reaches $2.0\times10^{43}$  (HY)!
The luminosity of Lyman $\alpha$ emission, which is the cooling radiation 
by interstellar gas, reaches $2.0\times10^{43}$ erg s$^{-1}$ at $t_{\rm age} = 0.1$ Gyr 
and $1.6\times10^{43}$ erg s$^{-1}$ at $t_{\rm age}=0.3$ Gyr,
respectively. They nicely match the observed luminosity of LAEs
and also well resemble LAEs with respect to other properties. 
After $t_{\rm age}\leq 0.5$ Gyr, the Lyman $\alpha$ luminosity quickly declines 
to several $10^{41}$ erg s$^{-1}$ that is lower than the detection limit. 
Then, the galaxy shifts to a stellar continuum radiation-dominated phase, 
which appears like LBGs \citep[see][]{MU06}.

\subsection{Escape Fractions}

\begin{figure}
\begin{center}
\includegraphics[scale=0.4]{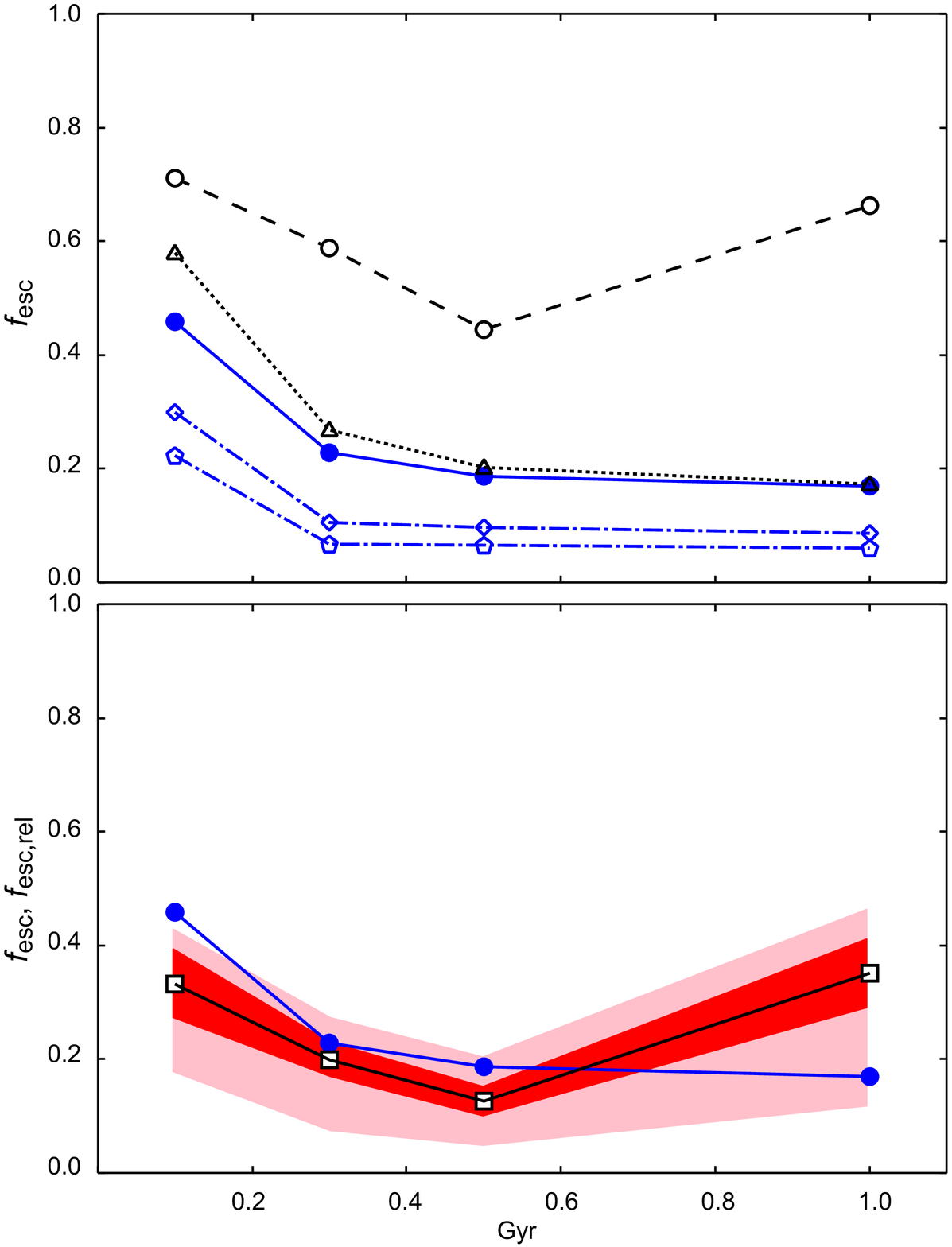}
\caption{
%--------
%$Upper ~panel:$
%Escape fractions $f_{\rm esc}$ for the simulated galaxy 
%as a function of \textcolor{red}{evolutionary time}. Filled circles 
%represent the escape fractions $f_{\rm esc}$ 
%with dust extinction, while 
%open circles denote $f_{\rm esc}$ without dust extinction. 
%The open triangles show the dust sputtering model.
%$Lower ~panel:$
%Comparison of $f_{\rm esc}$ with relative escape fractions 
%$f_{\rm esc, rel}$ that are derived from 
%the flux ratio at 1500~$\rm \AA$ to 900~$\rm \AA$
%[see equation (\ref{f_esc_rel})]. 
%The filled circles represent the absolute escape fractions 
%$f_{\rm esc}$ and the open squares do the relative escape 
%fractions $f_{\rm esc, rel}$. The variations by viewing angles
%are denoted by a pink belt with the standard deviation shown 
%by a narrower red belt.
%--------
%\textcolor{red}{
$Upper ~panel:$
Absolute escape fractions $f_{\rm esc}$ for the simulated galaxy 
as a function of evolutionary time.  
Blue symbols represent the escape fractions $f_{\rm esc}$ 
with dust extinction, where
blue filled circles, open diamonds, and open pentagons show 
the escape fraction for dust size of $\rm{0.1-1.0\mu m}$, 
$\rm{0.02-1.0\mu m}$, and $\rm{0.03-0.3\mu m}$, respectively.
Black open circles denote $f_{\rm esc}$ without dust extinction. 
The black open triangles show the dust sputtering model.
$Lower ~panel:$
Absolute escape fractions $f_{\rm esc}$ 
for dust size of $\rm{0.1-1.0\mu m}$
are compared to relative escape fractions $f_{\rm esc, rel}$ 
that are derived from the flux ratio at 1500~$\rm \AA$ to 900~$\rm \AA$
[see equation (\ref{f_esc_rel})]. 
The filled circles represent the absolute escape fractions 
$f_{\rm esc}$ and the open squares do the relative escape 
fractions $f_{\rm esc, rel}$. 
The variations by viewing angles are represented by a pink belt 
with the standard deviation shown by a narrower red belt.
%}
}
\end{center}
\end{figure}

Figure 2 shows the time evolution of escape fractions of ionizing photons
derived by the full radiation transfer calculations. 
The upper panel represents the absolute escape fractions, 
which are defined by
\begin{equation}
f_{\rm esc} \equiv \frac{N_{\rm esc}^{\gamma }}
{N_{\rm total}^{\gamma }},
\end{equation}
where $N_{\rm total}^{\gamma}$  
is the total number of pristine photons radiated from stars
and $N_{\rm esc}^{\gamma}$ is the number of photons 
escaped from the simulation box.
Filled circles in the upper panel of Fig. 2 represent the resultant
escape fractions $f_{\rm esc}$ with dust extinction.
In the LAE phase, $f_{\rm esc}=0.47$ at $t_{\rm age}=0.1$ Gyr and
$f_{\rm esc}=0.23$ at $t_{\rm age}=0.3$ Gyr, while 
in the LBG phase, $f_{\rm esc}=0.19$ at $t_{\rm age}=0.5$ Gyr and
$f_{\rm esc}=0.17$ at $t_{\rm age}=1.0$ Gyr. 
%\textcolor{blue}{
The escape fractions have the dependence on the size 
distributions of dust, since smaller grains result in lager extinction
for a given dust-to-gas ratio. 
If we extend the range to smaller grains as $\rm{0.02-1.0\mu m}$, 
the escape fractions are reduced to 
$f_{\rm esc} = 0.1-0.3$ in the LAE phase
and $f_{\rm esc} = 0.09-0.1$ in the LBG phase.
If the grains are even smaller as $\rm{0.03-0.3 \mu m}$, 
the escape fractions become 
$f_{\rm esc} = 0.07-0.22$ in the LAE phase
and $f_{\rm esc} = 0.06-0.07$ in the LBG phase.
These values are concordant with the estimations for LBGs 
by \citet{Sha06} and \citet{Iwa09}.
%}

%\textcolor{blue}{
The escape fractions are significantly regulated by interstellar dust.
In Fig. 2, we also show the dust-free model (open circles), 
where the absorption by interstellar dust is artificially switched off. 
We find that the dust extinction reduces $f_{\rm esc}$
by a factor of {1.5-8.5} in the LAE phase. On the other hand, 
The reduction by dust extinction is a factor of $2.5-11$ in the LBG phase. 
The change of the reduction factor is linked to
the enrichment of heavy elements in the galaxy.
%}

%\textcolor{red}{
\citet{Gne08a} have shown that the escape fractions are as small as
a few per cent, which is much smaller than the present results.
In the calculations by \citet{Gne08a}, most of young stars are distributed 
in the dense region of galactic disk. 
In our model, stars are extendedly distributed, and also 
the bulk of interstellar gas is collisionally ionized by the SN shock heating.
Since the interstellar medium is moderately optically-thin for ionizing photons, 
ionizing photons can escape through collisionally-ionized regions. 
%}
To assess the effects of collisional ionization, we have tentatively
calculated a case assuming $T=10^4$K and no collisional ionization.
As a result, we have found that
the escape fractions are reduced to be
$f_{\rm esc}=0.21$ at $t_{\rm age}=0.1$ Gyr and
$f_{\rm esc}=0.13$ at $t_{\rm age}=1.0$ Gyr. 
%\textcolor{red}{
This implies that the collisional ionization by shocks 
can contribute to enhance the escape fractions by a factor of two.
%}

%----------------------------------------------------------
%HY(new)!
%Furthermore, we consider that the interstellar dust may be vanished by 
%sputtering in the very high temperature gas ($T \gseq 10^6 K$). 
If dust grains are in high-temperature regions 
with $T \gseq 10^6 K$, they may be destructed by sputtering process.
The grain radius $a_{\rm d}$ decreases at a rate 
$da_{\rm d}/dt = - n_{\rm p}h_{\rm w}[1+(T_{\rm s}/T)^{2.5}]^{-1}$ cm s$^{-1}$
\citep{DS79,TM95,MB03}, where $h_{\rm w} = 3.2 \times 10^{-18} $ cm$^{4}$ s$^{-1}$
and $T_{\rm s} = 2 \times 10^{6}$ K are fitting parameters, 
and $n_{\rm p}$ is the proton number density.
Therefore, the destruction time scale of $\sim 0.1$ $\rm \mu$ m dust is 
shorter than the time scale of galaxy evolution ($\lseq 1$Gyr), 
if $n_{\rm p} \gseq10^{-4}$ cm$^{-3}$ and $T \gseq 10^{6}$ K.
Since the present model galaxy has $n_{\rm p} \gseq 10^{-4}$ on average
in high-temperature regions, 
we roughly suppose that all dust in high-temperature regions 
($T \gseq 10^6 K$) is evaporated by sputtering. 
Open triangles in the upper panel of Fig. 2 show 
the escape fractions when taking the sputtering into account.
After including dust sputtering, $f_{\rm esc}=0.58$ at $t_{\rm age}=0.1$ Gyr, and
$f_{\rm esc}=0.17$ at $t_{\rm age}=1.0$ Gyr.
At higher redshifts, the dust sputtering somewhat enlarges the escape fractions,
because a part of interstellar dust is distributed in high-temperature regions.
On the other hand, at lower redshifts, 
the results are basically the same as no-sputtering cases, since
ionizing photons are mainly absorbed by dust in low-temperature 
star-forming regions. 
Although the sputtering model here may overestimate the destruction of dust,
the escape fractions do not change very much 
by including the dust sputtering.
%HY(revised)
%\textcolor{red}
%{
Also, in our simulations,
the dust temperature does not rise over 100 K by photo-heating. 
Since the evaporation temperature of silicate dust is $\rm \sim10^{3} K$,
the photodestruction of dust is unimportant. 
%}

In order to evaluate the relative escape fractions defined by 
(\ref{f_esc_rel}), we make a mock observation of the 
simulated galaxy using the flux ratio at 1500~$\rm \AA$ to 900~$\rm \AA$. 
Assuming $(L1500/L900)_{\rm int} = 3$ \citep{2St01} and 
$Q(1500~\rm \AA)=1$, and also that 
the radiation flux at 1500~$\rm \AA$ is absorbed only by  
interstellar dust, we compute the transport of the radiation fluxes 
at 900~$\rm \AA$ and 1500~$\rm \AA$. The resultant relative escape 
fractions $f_{\rm esc, rel}$ are shown by open squares in the lower panel 
of Fig. 2. It turns out that
the predicted relative escape fractions ($f_{\rm esc, rel} \sim 0.1 - 0.3$) 
match the average of observed relative escape fractions as 0.14 \citep{Sha06}. 
%Deleted Iwata et al. (2008)
Also, it is consistent with the observational estimates given by 
\citet{2St01}, \citet{Gi02}, and \citet{In05,In06}.
%.....................................................................
%
%The efficient absorption by the interstellar dust makes the 
%radiation flux at 1500 ~$\rm \AA$ decrease. Accordingly, 
%the theoretical relative escape fraction ($f_{\rm esc, rel} 
%\sim 40\% - 100\% $) exhibits always greater than the intrinsic 
%escape fraction These $f_{\rm esc, rel}$ perfectly match the observed 
%average relative escape fraction 14 \% and $ > 15 \%$ reported by 
%\citet{Sha06} and \citet{Iwa09}, respectively. 
%($f_{\rm esc} = 17-47 \%$). 
%
%.....................................................................

%----------------------------------------------------------------------
%
% Section 4:  Discussion
%
%----------------------------------------------------------------------
\section{Discussion}

\citet{Sha06} estimated $f_{\rm esc, rel}$ for 14 LBGs in the SSA22a field,
and they detected ionizing photons from only 2 LBGs, that is, C49 and D3.
The reported relative escape fractions of 2 LBGs are extremely high as 
$f_{\rm esc,rel}({\rm C49}) = 0.65$, and $f_{\rm esc,rel}({\rm D3}) \geq 1.0$.
For the other 12 LBGs, only upper limits are suggested. 
\citet{Iwa09} reported the results of Subaru/Suprime-Cam deep imaging 
observations of the same field. They detected ionizing radiation from 7 LBGs 
as well as from 10 LAE candidates. They also showed the large scatter of 
the observed UV to Lyman continuum flux density ratios. For the seven 
detected LBGs, the ratio ranges from 2.4 to 23.8 with a median value of 6.6. 
%\textcolor{red}{
Then, the relative escape fractions for 7 LBGs ranges from 0.03 to 0.30 
after a correction for dust extinction, and
can be $0.05-0.55$ if IGM extinction is taken into account.
%}
In addition, some of the detected galaxies show significant spatial offsets 
of ionizing radiation from non-ionizing UV emission. 

The spatial distributions of interstellar dust are highly inhomogeneous 
in LAEs, because it should follow the inhomogeneous heavy-element distributions 
\citep[see also][]{MFM02,MUF04}. Therefore, the observed escape fractions 
strongly depend on viewing angles. 
We compute the probability distribution function, 
$ dp(f) / df \equiv N(f) / N_{\rm total}$, 
as a function of relative escape fraction along a line of sight, where 
$N_{\rm total}=128^2$ is the total number of bins in viewing angles and
$N(f)$ is the number of angular bins with a given escape fraction $f$.
Figure 3 shows the probability distribution function from redshift $t_{\rm age}=0.1$ to 1.0 Gyr. 
% MM
%Each line is $dp(f) / df$ at $z=7.0$ (green), 
%$5.8$ (blue), $5.0$ (magenta), and $3.7$ (red), respectively. 
%
This figure clearly shows that the relative escape fractions vary by 
viewing angles significantly.
For example, depending on viewing angles, the relative escape fractions 
$f_{\rm esc, rel}$ vary from 0.18 to 0.43 at $t_{\rm age}=0.1$ Gyr and from 0.12 to 
0.47 at $t_{\rm age}=1.0$ Gyr.
These probability distributions are indicated by a pink belt
in the lower panel of Fig. 2. 
The standard deviation at each redshift is also shown by 
a red belt.
This effect allows some of observed LAEs and LBGs to have 
as high escape fraction as reported by \citet{Iwa09}.

\begin{figure}
\begin{center}
\includegraphics[scale=0.4]{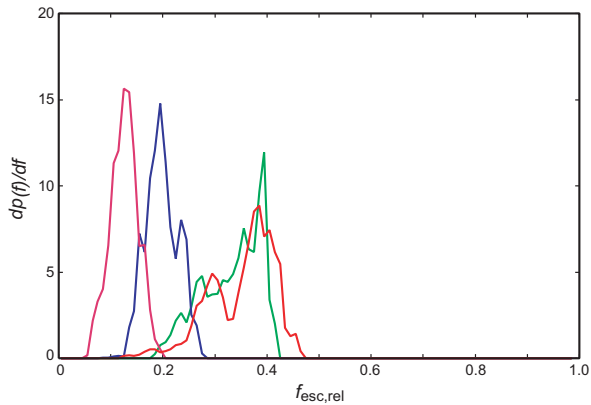}
\caption{
Probability distribution of relative escape fractions. 
% MM
Each line corresponds to the probability distribution at $t_{\rm age}=0.1$Gyr (green), 
$0.3$ Gyr (blue), $0.3$Gyr (magenta) and $1.0$ Gyr (red), respectively.
%Each line corresponds to the probability distribution at $z=7.0$ (magenta), 
%$5.8$ (blue), $5.0$ (green) and $3.7$ (red), respectively.
%
}
\end{center}
\end{figure}

The present analyses show the absolute escape fractions $f_{\rm esc}$ 
for LAEs and LBGs can be as large as $f_{\rm esc} \gseq 0.17$. 
Hence, LAEs and LBGs are potential sources for the IGM ionization
at $z\gseq 4$.
%\textcolor{red}{
Here, we quantify the contributions of LAEs and LBGs 
to the IGM ionization.
Specifically, we assess the emission rate of ionizing photons 
from LAEs and LBGs per unit comoving volume.
The emission rate is evaluated from the star formation rate 
based on the observed luminosity functions, with
coupling the escape fractions for LAE and LBG phases
obtained in the present analysis.
We use the average escape fraction $<f_{\rm esc}>=0.35$ for 
the LAE phase and $<f_{\rm esc}>=0.18$ for the LBG phase.
%}

In Figure 4, the emission rate of ionizing photons 
per comoving $\rm{Mpc^3}$ is shown as a function of redshift.
Open symbols depict the emission rate for 
the samples of LAEs that listed in the caption. 
Filled symbols show the emission rate for LBGs, which is estimated 
by extrapolating the luminosity function to $L=0.1L_{z=3}^{*}$ from \citet{St99} 
with dust extinction of E(B-V)=0.13 \citep[see also][]{Yo06}.
As a theoretical criterion, we adopt that by \citet{MHR99}, where
the emission rate of ionizing photons required to balance 
the recombination is given by
%\begin{equation}
%\dot{N}_{\rm ion} = 10^{51.2} \left( \frac{C}{30} \right) \left( \frac{1+z}{6}\right)^3 
%{~~\rm s^{-1} Mpc^{-3}},
%\end{equation}
$\dot{N}_{\rm ion} = 10^{47.4} C (1+z)^3 {~~\rm s^{-1} Mpc^{-3}}$.
$C$ is the clumping factor which parameterizes the inhomogeneity 
of ionized hydrogen in the IGM. \citet{2MHR99}
adopted $C=30$, based on the value computed by a cosmological radiative 
transfer simulation of \citet{GO97}. \citet{Ou04} 
also adopted $C=30$ when computing the number of ionizing photons needed 
to keep the IGM highly ionized at $z=5$, as do a number of other authors. 
Although $C$ may have some uncertainty, we assume $C=30$ here
as a fiducial value.

%----------------------------------------------------------
%HY
%The red broken lines in Fig. 4 show the emission rate of ionizing 
%photons per comoving Mpc as a function of redshift, which are derived 
%from the comoving star formation rate per unit volume in LAEs inferred 
%by Ouchi et al. (2003) for $z=3.1, 3.7$ and 5.7, Taniguchi et al. (2005) 
%for $z=6.6$, and Iye et al. (2006) for $z=7$. We used $<f_{\rm esc}>=35 \%$ 
%and $<f_{\rm esc}>=100 \%$ for the lower line and the upper line, 
%respectively.
%The blue broken lines in Fig. 4 also show that in LBGs inferred by 
%Sawicki \& Thompson et al. (2003) for $z=3 {\rm and} 4$, 
%Ouchi et al. (2003) for $z=5$, and Bunker et al. (2006) for $z=6$. 
%We used the average escape fraction $<f_{\rm esc}>=18 \%$ and 
%$<f_{\rm esc}>=100 \%$ for the lower line and the upper line, respectively.

%----------------------------------------------------------
%HY(new)!
Figure 4 clearly illustrates that observed LBGs can provide the majority 
of ionizing photons at $z=3-5$, and play an important role 
to keep the universe ionized.
However, LAEs are not capable of ionizing large volumes at the redshift $z=3-5$.
On the other hand, ionizing photons from not only LBGs but also
LAEs are not enough to ionize the IGM at $z \gseq 6$.
Most of photons that ionize the universe 
may come from undetected faint LAEs and LBGs or other sources.
Recently, \citet{CF07} studied the cosmic reionization history to 
account for a number of observational data, and pointed out that 
low-mass galaxies hosting Pop III stars can be predominant ionizing sources 
of the IGM at high-$z$. Our results advocate their model.
In the present analysis, all Lyman $\alpha$ photons are assumed to escape. 
Hence, the contribution of LAEs may be underestimated. 
In the future work, we intend to include Lyman $\alpha$ line transfer 
to compare the numerical results more precisely with the observations.%----------------------------------------------------------

\begin{figure}
\begin{center}
\includegraphics[scale=0.4]{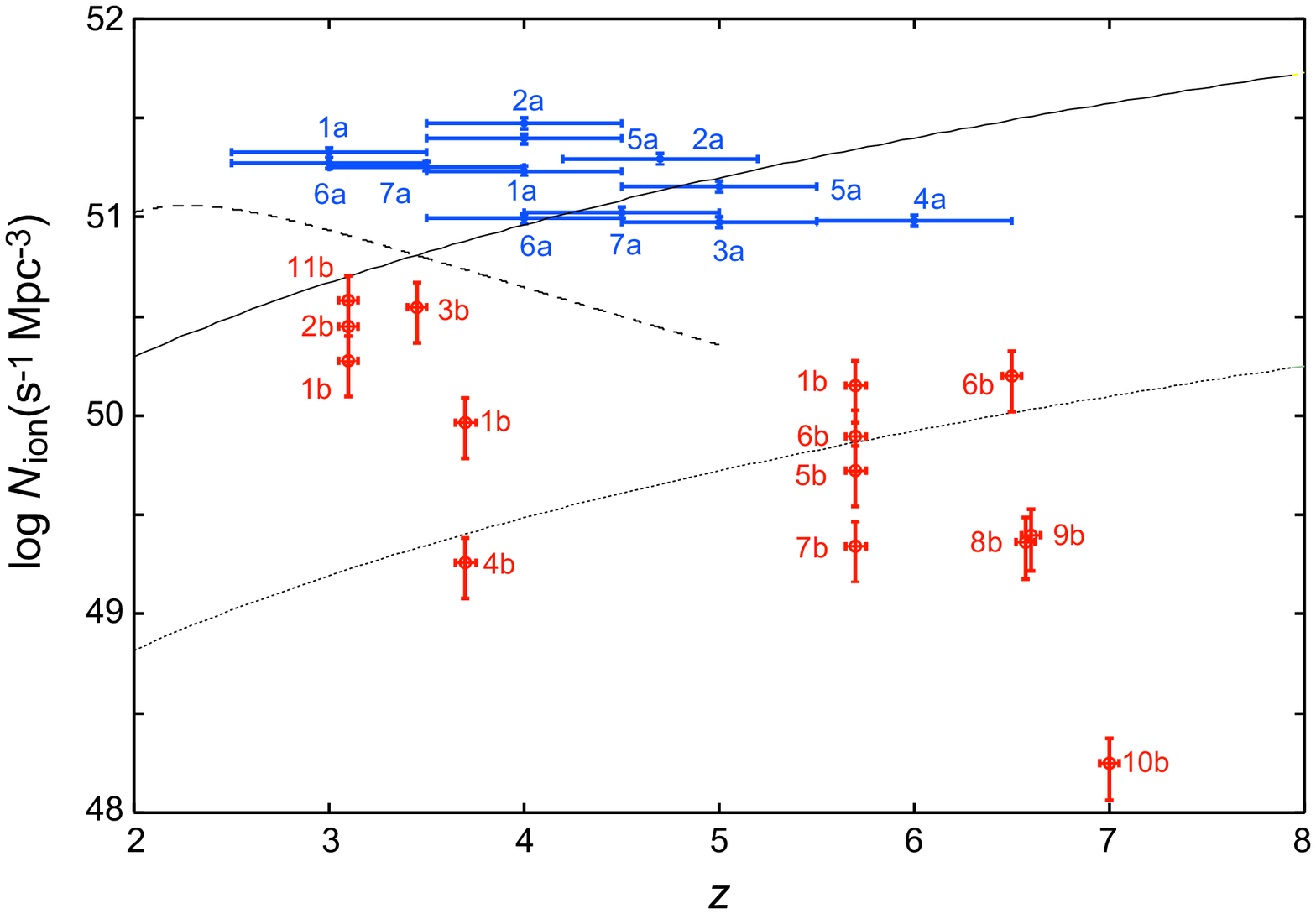}
\caption{
 The evolution of emission rate of ionizing photons per comoving Mpc, 
$\dot{N}_{\rm ion}$, as a function of redshift.
Open symbols represent $\dot{N}_{\rm ion}$ of LAEs derived from 
1a) \citet{Ou08},  
2a) \citet{Ku00}, 
3a) \citet{Br05}, 
4a) \citet{Fu03}, 
5a) \citet{Aj03}, 
6a) \citet{MR04},  
7a) \citet{Rh03},
8a) \citet{Ko03}, 
9a) \citet{Ta05}, 
10a) \citet{Iye06},
and 
11a) \citet{Gr07} 
with $<f_{\rm esc}>=0.35$ which is mean escape fraction at LAE phase. 
The blue filled symbols represent the $\dot{N}_{\rm ion}$ of LBGs derived from
%1b) Steidel et al. (2001), 
%2b) Yoshida et al. (2006),
%3b) Iwata et al. (2003), 
%4b) Bowens et al. (2006),
%5b) Ouchi et al. (2008), 
%6b) Sawicki\& Thompson (2006),
%and 
%7b) Gabash et al. (2004) 
1b) \citet{St99}, 
2b) \citet{Yo06},
3b) \citet{Iwa03}, 
4b) \citet{Bou06},
5b) \citet{Ou04}, 
6b) \citet{ST06},
and 
7b) \citet{Gab04} 
with $<f_{\rm esc}>=0.18$ which is mean escape fraction at LBG phase.
 The horizontal and vertical error-bars are arisen from the uncertainty of 
observations and the variation of escape fractions 
(LAE : $f_{\rm esc} = 0.22 - 0.47$, LBG : $f_{\rm esc} = 0.17 - 0.19$), respectively.
A solid line and a dotted line indicate the emission rate 
required to ionize the IGM with $C=30$ and $C=1$, respectively
\citep{2MHR99}.
A dashed line represents the emission rate evaluated by the QSO luminosity function.
}
\end{center}
\end{figure}

\section{SUMMARY}
We have performed three-dimensional radiation transfer calculations, 
based on a high-resolution hydrodynamic simulation of a supernova-dominated
primordial galaxy, to obtain the ionization structure and explore 
the escape fractions of ionizing photons from LAEs and LBGs at high redshifts.
%\textcolor{blue}{
The effect of dust extinction is incorporated according to the chemical enrichment, 
taking the size distributions of dust into account. 
As a result, we find that dust extinction reduces the escape fractions 
by a factor of $1.5-8.5$
in the LAE phase and by a factor of $2.5-11$ in the LBG phase.
The resultant escape fractions are $0.07-0.47$ in the LAE phase and
$0.06-0.17$ in the LBG phase. 
These results are well concordant with recent observations. 
We have found that the combination of diffuse distributions of stars
and supernova shock-heating is important for UV escape fractions. 
In the present galaxy model, young stars are extendedly distributed, and also 
the bulk of interstellar gas is collisionally ionized by supernova shock-heating. 
Since the interstellar medium is moderately optically-thin and
quite bubbly, ionizing photons can escape through collisionally-ionized regions. 
The collisional ionization by shocks contributes 
by a factor of $\approx 2$ to the increase of the escape fractions.
%}

The relative escape fractions derived by mock observations
of the simulated galaxy match quite well the estimates by
recent observations for LAEs and LBGs.
To assess the contribution of LAEs and LBGs
to the IGM ionization, the resultant escape fractions have been combined 
with the luminosity functions of LAEs and LBGs.
As a result, we find that high-$z$ LAEs and LBGs can 
ionize the IGM at $z=3-5$. 
However, ionizing radiation from LAEs as well as LBGs is not enough to ionize 
the IGM at $z>6$. That implies that undetected faint LAEs and LBGs or 
additional ionization sources may determine the IGM ionization at $z>6$. 

%%  MM
%\textcolor{magenta}{
Very recently, \citet{WC09} performed 3D radiation hydrodynamic simulations
to assess the contribution of dwarf galaxies to cosmic reionization 
at redshift $z=8$. They studied the UV escape fractions for 
low-mass galaxies in the mass range of $M_{\rm total}=3\times10^{6}-3\times10^{9}M_{\odot}$.
As a result, they have shown that the UV escape fractions can reach
up to $\sim$ 0.8 without dust extinction in halos with $>10^8M_\odot$
for a top-heavy initial mass function (IMF).
However, in order to assess the contribution of LAEs and LBGs to IGM reionization,
we should evaluate the UV escape fractions from more massive galaxies with dust extinction.
Here, we have shown that a high-mass, metal-enriched galaxy
at low redshifts can allow escape fractions as large as some tens per cent .
We have found that an order-of-magnitude increase in the escape fractions can be attributed to the diffuse distribution of stars, while the collisional ionization further raises them by some factor.
%}

%%%HY(revised)
%%\textcolor{red}{
%%So far, some numerical simulations have shown that 
%%large escape fractions of some tens per cent are possible for
%%high redshift ($z \gseq 8$) low-mass galaxies, which include Pop III stars
%%\citep{ABS06, Wh04, Ki04, WC09}.
%%Other simulations of high-mass galaxies at low redshifts ($z \sim 3$) 
%%indicate small escape fractions of $\sim 3 $ per cent \citep{Raz06, Raz07, Gne08a}. 
%%Here, we have shown that a high-mass, metal-enriched galaxy
%%at low redshifts can allow escape fractions as large as some tens per cent .
%%We have found that the diffuse distributions of stars mainly enlarge
%%the escape fractions by an order of magnitude and the collisional 
%%ionization further enhances them by some factor.
%%}

%----------------------------------------------------------------------
%
% Acknowledge
%
%----------------------------------------------------------------------
\section*{Acknowledgments}
We are grateful to A. Inoue, A. Ferrara, and A. Razoumov
for valuable discussion and comments. 
Numerical simulations have been performed with the {\it FIRST} simulator  
and {\it T2K-Tsukuba} at Center for Computational Sciences, in University of Tsukuba. 
This work was supported in part by the {\it FIRST} project based on
Grants-in-Aid for Specially Promoted Research by 
MEXT (16002003) and
%Grant-in-Aid for Scientific Research (S) by
JSPS Grant-in-Aid for Scientific Research (S) (20224002),
(A) (21244013), and (C) (18540242).

%----------------------------------------------------------------------
%
% References
%
%----------------------------------------------------------------------

\label{lastpage}

\end{document}